# Deep Sub-Wavelength Plasmon Lasers

Rupert F Oulton[1*], Volker J Sorger[1*], Thomas Zentgraf[1*], Renmin Ma[2], Christopher Gladden[1], Lun Dai[2], Guy Bartal[1] and Xiang Zhang[1,3]

[1]NSF Nanoscale Science and Engineering Centre, 3112 Etcheverry Hall, University of California, Berkeley, CA 94720, USA.

[2]State Key Lab for Mesoscopic Physics and School of Physics, Peking University, Beijing 100871, China

[3]Materials Sciences Division, Lawrence Berkeley National Laboratory, 1 Cyclotron Road, Berkeley, CA 94720, USA

*These authors contributed equally to this work.

**Laser science has tackled physical limitations to achieve higher power, faster and smaller light sources [1-8]. The quest for ultra-compact laser that can directly generate coherent optical fields at the nano-scale, far beyond the diffraction limit of light, remains a key fundamental challenge [9, 10]. Microscopic lasers based on photonic crystals [3], micro-disks [4], metal clad cavities [5] and nanowires [6-8] can now reach the diffraction limit, which restricts both the optical mode size and physical device dimension to be larger than half a wavelength. While surface plasmons [11-13] are capable of tightly localizing light, ohmic loss at optical frequencies has inhibited the realization of truly nano-scale lasers [14, 15]. Recent theory has proposed a way to significantly reduce plasmonic loss while maintaining ultra-small modes by using a hybrid plasmonic waveguide [16]. Using this approach, we report an experimental demonstration of nano-scale plasmonic lasers producing optical modes 100 times smaller than the diffraction limit, utilizing a high gain Cadmium Sulphide semiconductor nanowire atop a Silver surface separated by a 5 nm thick insulating gap. Direct measurements of emission lifetime reveal a broad-band enhancement of the nanowire's spontaneous emission rate by up to 6 times due to the strong mode confinement [17] and the signature of apparently threshold-less lasing. Since plasmonic modes have no cut-off, we show down-scaling of the lateral dimensions of both device and optical mode. As these optical coherent sources approaches molecular and electronics length scales, plasmonic lasers offer the**

**possibility to explore extreme interactions between light and matter, opening new avenues in active photonic circuits [18], bio-sensing [19] and quantum information technology [20].**

Surface plasmon polariton (SPP) are the key to breaking down the diffraction limit of conventional optics as they allow the compact storage of optical energy in electron oscillations at the interfaces of metals and dielectrics [11-13]. Accessing sub-wavelength optical length scales introduces the prospect of compact optical devices with new functionalities as it enhances inherently weak physical processes, such as fluorescence and Raman scattering of single molecules [19] and non-linear phenomena [21]. An optical source that couples electronic transitions directly to strongly localized optical modes is highly desirable as it would avoid the limitations of delivering light from a macroscopic external source to the nano-scale, such as low coupling efficiency and difficulties in accessing individual optical modes [22].

Achieving stimulated amplification of SPPs at visible frequencies remains a challenge due to the intrinsic ohmic losses of metals. This has driven recent research to examine stimulated SPP emission in systems that exhibit low loss, but only minimal confinement, which excludes such schemes from the rich new physics of nanometre scale optics [14, 15]. Recently, we have theoretically proposed a new approach hybridizing dielectric waveguiding with plasmonics, where a semiconductor nanowire atop a metallic surface, separated by a nano-scale insulating gap [16]. The coupling between the plasmonic and waveguide modes across the gap enables energy storage in non-metallic regions. This hybridization allows SPPs to travel over large distances with strong mode confinement [23] and the integration of a high quality semiconductor gain material with the confinement mechanism itself.

In this Letter, we utilize the "hybrid plasmonics" approach to experimentally show the laser action of SPPs with mode areas as small as $\lambda^2/400$. The truly nano-scale plasmonic laser devices consist Cadmium Sulphide (CdS) nanowires [24] on a silver film, where the gap layer is Magnesium Fluoride ($MgF_2$, Fig 1a). The close proximity of the semiconductor and metal interfaces concentrates light into an extremely small area as much as 100 times less than the diffraction limit [16] (Fig 1b). To show the unique properties of hybridized plasmon modes, we compare the plasmonic lasers directly with CdS nanowire lasers on a quartz substrate, similar to typical nanowire lasers reported before [6-8]. In what follows, we will refer to these two devices as *plasmonic* and *photonic* lasers respectively.

We optically pump these laser devices at a wavelength of 405 nm and measure emission from the dominant $I_2$ CdS exciton line at 489 nm [25]. At moderate pump intensities (0.1-0.5 kWcm$^{-2}$), we observe the onset of amplified spontaneous emission peaks. These correspond to the longitudinal cavity modes that form when plasmonic propagation losses are compensated by gain allowing plasmonic modes to resonate between the reflective nanowire end-facets (Fig 2a). The clear signature of multiple cavity mode resonances at higher pump powers demonstrates sufficient material gain to achieve full laser oscillation, shown by the non-linear response of the integrated output power with increasing input intensity (Fig 2a inset).

Surmounting the limitations of conventional optics, plasmonic lasers not only support nano-scale optical modes, their physical size can also be much smaller than conventional lasers, i.e., plasmonic lasers operate under conditions where photonic modes cannot even exist [8]. Plasmonic lasers maintain strong confinement and optical mode-gain overlap over a broad range of nanowire diameters (inset i and ii of Fig 2b) with only a weak dependence on the nanowire diameter. While hybrid modes do not experience mode cutoff, the threshold intensity increases for smaller nanowires due to the reduction in the total gain volume. Conversely, photonic lasers exhibit a strong dependence of the mode confinement on the nanowire diameter (inset iii and iv of Fig 2b), resulting in a sharp increase in the threshold intensity at diameters near 150 nm due to a poor overlap between the photonic mode and the gain material. Moreover, actual photonic lasers suffer mode *cut-off* as the leakage into the quartz substrate prevents lasing for nanowire diameters less than 140 nm [8]. This is to say, the observation of plasmonic lasing for nanowire diametres of just 52 nm confirms the role of the hybrid plasmonic mode where a purely dielectric nanowire mode cannot exist.

Plasmonic modes often exhibit highly polarised behaviour as the electric field normal to the metal surface binds most strongly to electronic surface charge. We have detected the signature of lasing plasmons from the polarisation of scattered light from the nanowire end-facets, which is in the same direction as the nanowire. Conversely, the polarisation of scattered light from photonic lasers is perpendicular to the nanowire, This distinction provides a direct confirmation of the plasmonic mode.

We find a strong increase of the spontaneous emission rate when the gap size between the nanowire and metal surface is decreased. Lifetime measurements reveal a Purcell factor of more than 6 at gap width of 5 nm and for nanowire diameters near 120 nm (Fig 3), where hybrid plasmonic modes are most strongly localized [16]. This enhancement factor corresponds to a mode that is 100 times smaller than the diffraction limit, which agrees well with mode size calculations. While the enhanced emission rate, or Purcell effect [17], is usually associated with high quality micro-cavities [3-5], we observe a broad-band Purcell effect arising from mode confinement alone without a cavity [20, 26].

We next examine the physics underlying the gain mechanism in the plasmonic lasers, which combines exciton dynamics, the modification of spontaneous emission [17] and mode competition. While photo-generated excitons have intrinsic lifetimes of up to 400 ps [25], they recombine faster at the edge of the nanowire near the gap region due to strong optical confinement mediated by the hybrid plasmon mode (Fig 3). The exciton diffusion length in bulk CdS is about a micometre [27], which is much larger than the nanowire diameter. Therefore, the distribution of exciton recombination quickly adjusts itself to match the hybrid mode's profile (see Fig 1b). The fast diffusion and the enhanced emission rate into the hybrid plasmonic mode lead to preferential plasmon generation. In this way, the proportion of light that couples into the laser mode, known as the *spontaneous emission factor, β* can be high (Fig 3b) [28]. The measured emission rates and a simple emission model show that the β-factor of the plasmonic mode is as high as

80% for a gap width of 5 nm (see Methods). For gap widths below 5 nm, the exciton recombination is too close to the metal surface, causing rapid non-radiative quenching to lossy surface waves [29]. While nanowires placed in direct contact with the metal surface show the highest spontaneous emission rates, these devices exhibit weak luminescence and do not lase. Our calculations support these observations indicating a sharp reduction in the β-factor below gap widths of 5 nm.

A high spontaneous emission factor is often associated with low threshold laser operation where undesired emission modes are suppressed. The laser threshold is commonly manifested as a *kink* between two linear regimes of the output power versus pump intensity response. However, it is known that lasers with strong mode confinement do not necessarily exhibit such behaviour so that the laser threshold may be obscured [28, 30]. Since the plasmonic lasers exhibit strong mode confinement, we indeed observe this *smearing* of the threshold pump intensity (Fig 4). The photonic lasers we measured, on the other hand, show the distinctive kink in the output power, which is in agreement with recently reported Zinc Oxide nanowire lasers [8]. Since both plasmonic and photonic lasers in this work utilize the same gain material, we conclude that the smeared response in output power arises from local electromagnetic confinement. We therefore attribute this distinct behaviour to the increased spontaneous emission factor of the hybrid plasmon mode. While the linear response, shown in Fig 4 for a nanowire diameter of 66 nm, may indicate that the laser is nearly threshold-less [28], in reality, the onset of amplified spontaneous emission only occurs once cavity losses are compensated.

The demonstration of deep sub-wavelength plasmon laser action at visible frequencies shows the way to new sources that produce coherent light directly at the nano-scale. The increase of the spontaneous emission rate by up to 6 times shows that extremely strong mode confinement and a high spontaneous emission factor, providing a route to deep sub-wavelength lasers based on plasmonics. Furthermore, we have shown that the advantage of plasmonic lasers is the ability to down-scale the physical size of devices, as well as the optical modes they contain, unlike diffraction limited lasers. The impact of plasmonic lasers on optoelectronics integration is potentially significant as the optical fields of these devices rival the smallest commercial transistor gate sizes and thereby reconcile the length scales of electronics and optics.

**Methods**

**Spectral and Lifetime Measurements**

A frequency doubled, mode-locked Ti-Sapphire laser (Spectra Physics) was used to pump the plasmonic and photonic lasers ($\lambda_{pump}$ = 405 nm, repetition rate = 80 MHz, pulse length = 100 fs). An objective lens (20X, NA = 0.4) was used to focus the pump beam to a 38 μm diameter spot on the sample. All experiments were carried out at low temperature, T < 10 K, utilizing a liquid-He cooled cryostat (Janis Research Company). Individual spectra were recorded using a spectrometer (Princeton Instruments) and a liquid-N2 cooled CCD. The lifetime measurements were conducted under very low pump conditions to avoid heating and exciton scattering effects using time-correlated single photon counting (Picoharp 300, Micro Photon Devices) (see supplementary online material). A 490 ± 10 nm band pass filter was used to filter out ambient light and pass light from the $I_2$ CdS exciton line.

**Plasmonic Laser Fabrication**

The technical challenge of constructing the plasmonic lasers is to ensure a good contact between the nanowire surface and the planar optical films. We have been able to control this gap width, h, to approximately 2 nm accuracy due to low film roughness (<2 nm) and deposition accuracy (~10%). The CdS nanowires do not contribute significantly to poor contact as they are practically single crystals exhibiting extremely low surface roughness (see supplementary online material). The nanowires were grown using Chemical Vapour Deposition and self assembly from a thin Au film seeding layer. Individual nanowire diameters, d, are therefore random, but are tuneable to the range of 50-500 nm. We have studied the interaction of nanowire emission with the Silver film using 5 different devices. CdS nanowires were deposited from solution by spin coating onto pre-prepared films with varying Magnesium Fluoride thicknesses of 0, 5, 10 and 25 nm, along with control devices of CdS nanowires on a quartz substrate.

**Simple emission model for calculating Purcell and β Factors**

The Purcell effect [17] modifies the emission rate, $\gamma(d,h)$, of the CdS nanowire such that $\gamma(d,h) = F(d,h)\gamma_0 = (1 + f_1 F_{SP}(d,h) + f_2 h^{-3})\gamma_0$. Here, $F(d,h)$ is the Purcell Factor and $F_{SP}(d,h)$ is the peak rate enhancement due to the hybrid plasmonic mode. In comparing the theoretical model with the experimental data, three free-parameters were determined, which are assumed not to depend on d and h. Here, $f_1 = 0.67$ accounts for the average SPP rate enhancement due to the distribution of excitons in the nanowire; $f_2 = 23.25\,nm^3$ accounts for the proportion of LSWs [28] and $\gamma_0 = 2.49$ ns$^{-1}$ is the intrinsic emission rate of I$_2$ excitons in CdS [25]. The spontaneous emission factor, $\beta(d,h)$, for the lasing mode follows from the numerical fit to the rate enhancement data. Defined as the ratio of the emission rate into the SPP mode and the total emission rate, $\beta(d,h) = f_1 F_{SP}(d,h) F(d,h)^{-1}$.

**Figures**

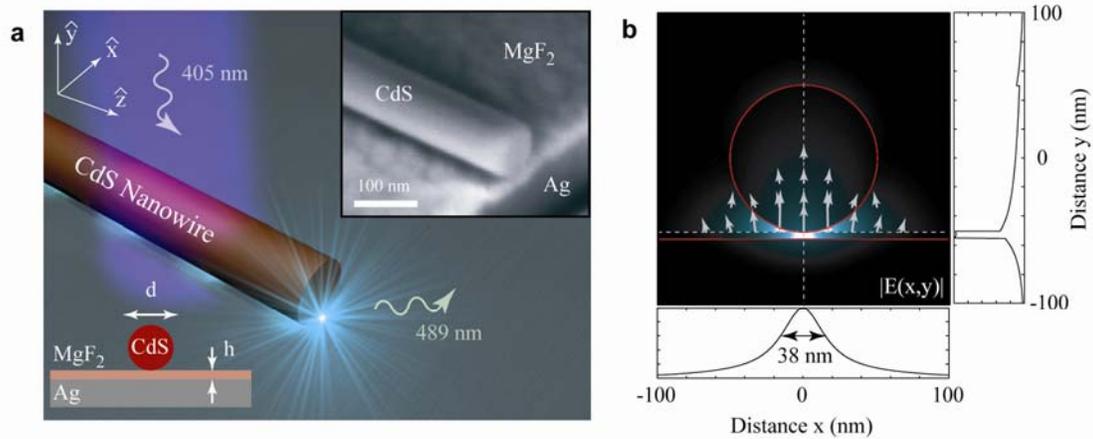

**Figure 1. The deep sub-wavelength plasmon laser. a** The plasmon laser consists of a Cadmium Sulphide semiconductor nanowire atop a Silver substrate, separated by a nanometre scale $MgF_2$ layer. This structure supports a new type of plasmonic mode [16] whose mode size can be 100 times smaller than a diffraction limited spot. The inset shows a scanning microscopy image of a typical plasmon laser, which has been sliced perpendicular to the nanowire's axis to show the underlying layers. Panel **b** shows the electric field distribution and direction $|E(x,y)|$ of a hybrid plasmonic mode at a wavelength $\lambda = 489$ nm corresponding to the CdS $I_2$ exciton line [25]. The cross-sectional field plots (along broken lines in field map) illustrate the strong overall confinement in the gap region between nanowire and metal surface with sufficient modal overlap in the semiconductor to facilitate gain.

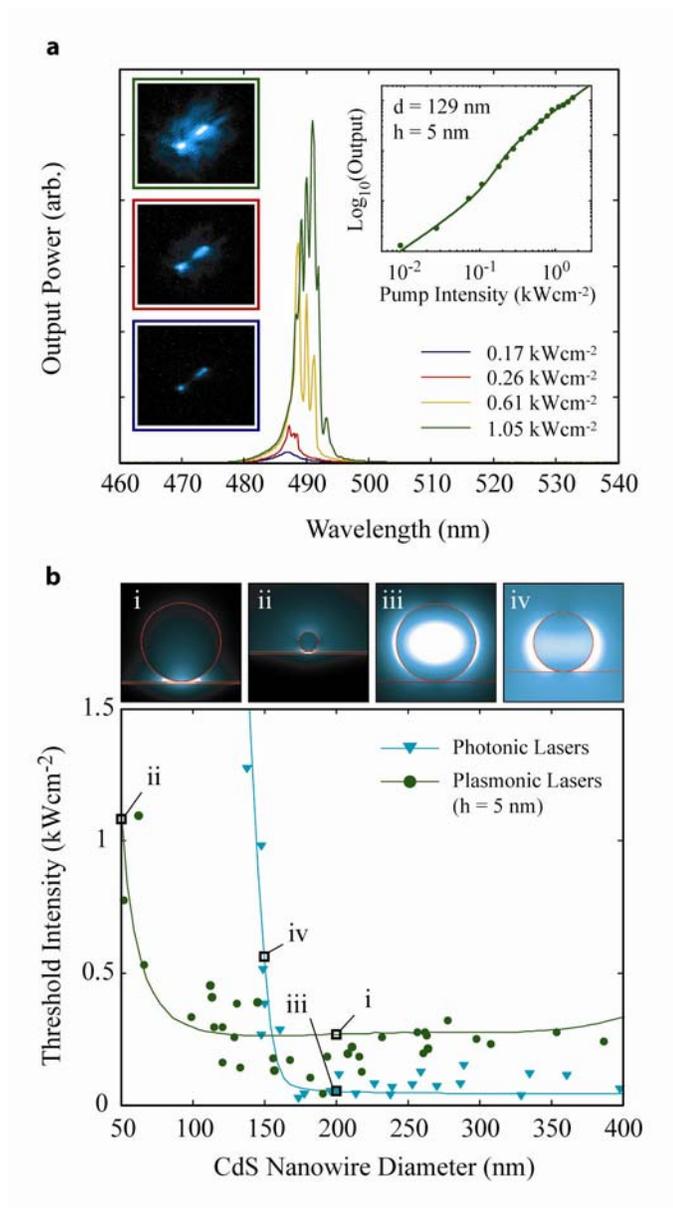

**Figure 2. Laser oscillation and threshold characteristics of plasmonic and photonic lasers.**
**a** Laser oscillation of the plasmonic laser (longitudinal modes). The four spectra for different pump intensities exemplify the transition from spontaneous emission (0.17 kWcm$^{-2}$) via amplified spontaneous emission (0.26-0.61 kWcm$^{-2}$) to full laser oscillation (>0.61 kWcm$^{-2}$). The top right inset shows the non-linear response of the output power to the pump intensity. The insets on the left show the corresponding microscope images of a plasmon laser with d = 66 nm exhibiting spontaneous emission, amplified spontaneous emission and laser oscillation, where the scattered light output is from the end-facets. **b** Threshold behaviour of plasmonic and photonic lasers with nanowire dimaetre. The experimental data points correspond to the onset of amplified spontaneous emission, which are considered to be the lasing threshold. Amplified spontaneous emission in hybrid plasmonic modes occurs at moderate pump intensities of 0.1-0.5 kWcm$^{-2}$ across a broad range of diameters. This is attributed to the insensitivity of the mode confinement to the nanowire diameter (insets i-ii) and the ability of the mode to remain confined even for a 50 nm diameter wire (inset ii). While the photonic lasers have similar threshold intensities around 0.1 kWcm$^{-2}$ for all nanowires larger than 200 nm (inset iii), a sharp increase in the threshold occurs for diameters near 150 nm, due to the loss of confinement within the nanowire and subsequent lack of overlap with the gain region [8] (inset iv). The solid lines show a best numerical fit to a simple rate equation model. Below 140 nm, the photonic mode is *cut-off* and lasing could not be observed at all. In contrast, plasmonic lasers maintain strong confinement and optical mode-gain overlap for diameters as small as 52 nm, a diameter for which a photonic mode does not even exist.

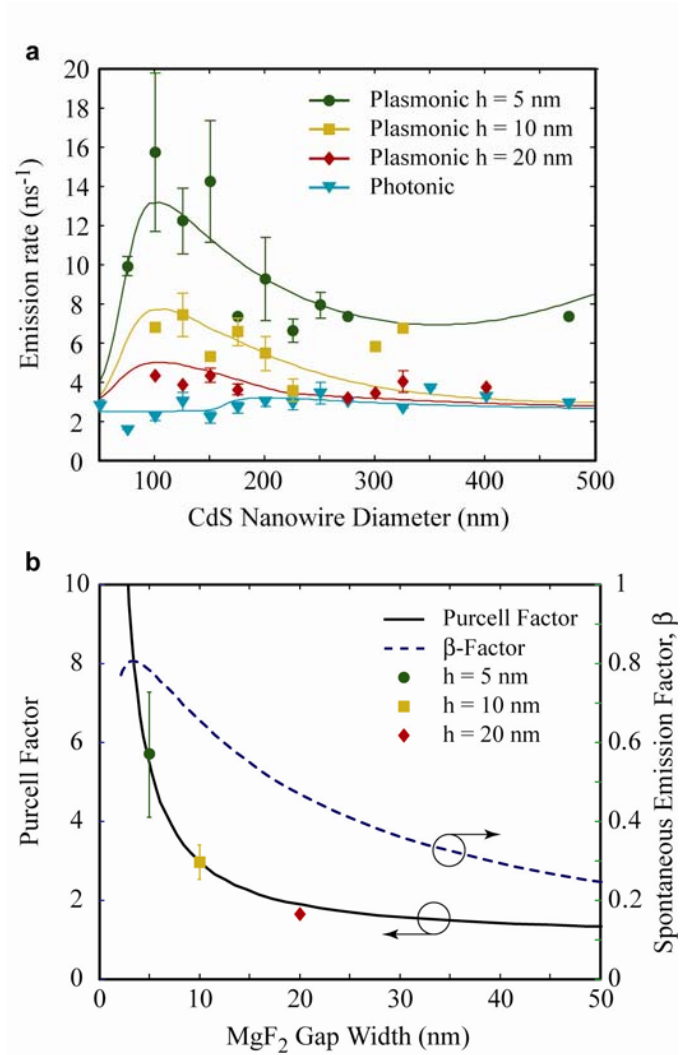

**Figure 3. The Purcell effect in plasmonic and photonic lasers.** Panel **a** shows the emission rates of photonic and plasmonic nanowire lasers with different $MgF_2$ gap widths as a function of the nanowire diameter following the calculated trend. The optimal confinement condition of hybrid plasmonic modes is found near d = 120 nm, where the hybridization of nanowire and SPP modes is strongest giving the highest emission rate [16]. The emission rates of nanowires placed in direct contact with the metal surface are given in the SOM. Panel **b** shows the Purcell factors determined from a numerical fit of the emission rate measurements to a simple emission model (see Methods). Near optimal confinement (d = 120 nm ± 20 nm), the average Purcell factor for devices with 5 nm gaps

is more than 6, which is considered high for a broad band effect. We have also calculated the β-factor from the emission model fit by accounting for the possible emission pathways (see Methods). The β-factor reaches a maximum of 80 % for a gap width near 5 nm. At gap widths smaller than 5 nm, non-radiative quenching to the metal surface causes a sharp drop in the hybrid plasmon β-factor, which subsequently eliminates the possibility for lasing.

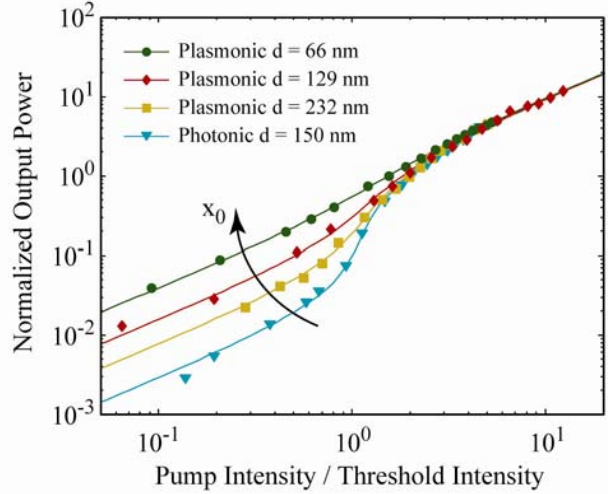

**Figure 4. Nearly *threshold-less lasing* due to high spontaneous emission factor.** The dependence of measured output power over pump intensity highlights clear differences in the physics underlying the plasmonic (h = 5 nm) and photonic lasers using a multi-mode lasing model [30]. In particular, our fitting parameter, $x_0$, is related to the saturation of gain into individual longitudinal laser modes and their lateral mode area. A higher value of $x_0$ corresponds to a smaller mode area and a higher β-factor. Photonic lasers exhibit a clear transition between spontaneous emission and laser operation characterized by a change in the gradient of input intensity versus output power, corresponding to the laser threshold. For a 150 nm diameter nanowire, the parameter $x_0 = 0.026$ [30] is in agreement with a recent nanowire laser study [8]. Plasmonic lasers, however, show a strong dependence of $x_0$ in the nanowire diameter; a large multi-mode plasmonic laser (d = 232 nm) shows a somewhat smeared transition region ($x_0 = 0.074$), while smaller single mode plasmonic lasers (d = 129 nm and d = 66 nm) have much less visible changes in gradient ($x_0 = 0.150$ and $x_0 = 0.380$ respectively). The large value of $x_0$ observed in the plasmonic lasers is associated with threshold-less operation and attributed to the strong mode confinement giving rise to a high spontaneous emission factor. Nevertheless, the onset of amplified spontaneous emission peaks in all devices occurred at non-zero threshold intensities as shown in Fig 2b.